# Ion irradiation induced direct damage to DNA

Wei WANG [*,†,‡]    Zengliang YU [‡]    Wenhui SU [†]

[†] Center for the Condensed Matter Science and Technology, Harbin Institute of Technology, Harbin 150080, China

[‡] Key Lab of Ion Beam Bioengineering, Institute of Plasma Physics, Chinese Academy of Sciences, P. O. Box 1126, Hefei, 230031, China

*Corresponding author. Email: wwang_ol@hit.edu.cn

**Abstract:** Ion beams have been widely applied in a few biological research fields such as radioactive breeding, health protection, and tumor therapy. Up to now many interesting and impressive achievements in biology and agriculture have been made. Over the past several decades, scientists in biology, physics, and chemistry have pursued investigations focused on understanding the mechanisms of these radiobiological effects of ion beams. From the chemical point of view, these effects are due to the ion irradiation induced biomolecular damage, direct or indirect. In this review, we will present a chemical overview of the direct effects of ion irradiation upon DNA and its components, based on a review of literature combined with recent experimental results. It is suggested that, under ion bombardment, a DNA molecule undergoes a variety of processes, including radical formation, atomic displacement, intramolecular bond-scissions, emission of fragments, fragment recombination and molecular crosslink, which may lead to genetic mutations or cell death. This may help to understand the mechanisms of the radiobiological effects caused by ion irradiation, such as radiation breeding and tumor therapy.

## 1. Introduction

Along with the development of neoteric accelerator techniques, there was a parallel worldwide growth in their practical application and now ion beams have been widely used in several applied research fields including radioactive breeding [1-3], health protection [4-7], and tumor therapy [8,9]. Since the late 1980s, low energy ion beam bioengineering research has been engaged, e.g., ion implantation to improve crops and microbes, ion beam mediated gene transfer [1]. Now it is being developed vigorously and impressive accomplishments in biology and agriculture have been achieved [1,2,10-12]. Meanwhile, ion-beam tumor therapy is under rapid development and has achieved significant clinical success, which benefits from the unique focused energy distribution of heavy ions at the "Bragg peak" [9,13,14].

Over the past several decades, scientists in chemistry, physics, and biology have pursued various investigations focused on understanding the mechanisms of the above-mentioned radiobiological effects of ion beams. Due to the complexity of ion-organism interaction, however, these phenomena heretofore cannot be unambiguously explained and the mechanisms of the biological effects remain unclear. What can be definitely confirmed is that those observable events are closely correlative to some chemical changes of biomolecules. Most of the previous research interests have focused on DNA which contains genetic information and can influence the phenotype of an organism. It has been suggested that the damage to the genome induced by sparsely ionizing radiation, such as X- and γ-rays, in a living cell by ionizing radiation is about two-thirds indirect and one-third direct [15,16]. The former, which has been reviewed extensively elsewhere [15,17], concerns radiation induced reactive species surrounding the biomolecules, such as radicals (e.g., •OH, H) and secondary electrons; the latter mainly results from direct collision-induced damage in the biomolecules. More recently, the assignment of the two effects has been improved. A simulation research reported that the yield of DNA strand breaks caused by ion irradiation corresponds to a bigger contribution of 40-45% direct effects, compared to 35% for γ irradiation [18].

In this contribution, therefore, the major emphasis is put on the direct ion-biomolecule interactions, which to the best of our knowledge has not been summarized. Herein we would refer the biomolecule as the fundamental material, DNA and its components. Detailed knowledge of reaction products and reaction mechanisms in the radiation chemistry of these molecules is of great importance in the development of our understanding of the precise chemical basis of radiation damage in biological systems.

## 2. Multiple interactions between ion irradiation and DNA
2.1. Nucleobases

The formation of free radicals in purine and pyrimidine crystals is a ubiquitous phenomenon when they are submitted to ion impact. As we all know, an incoming ion colliding with bulk matter is quickly neutralized upon interaction with the first atomic layers and produces an extremely heterogeneous chemical environment around the path of the individual particle, including the presence of a free radical rich environment. Within the model of ion-target collision, there are two kinds of interactions between the projectile and the target atom: nuclear collision and electron collision. In the case of electron collision, the ion transfers some of its energy to the electrons of the target atoms causing ionization, excitation, electron capture or loss, and then radical formation.

Deng and co-workers demonstrated that some ion impingement induced radicals have a long lifetime, even of more than one month [19]. This makes the use of electron spin resonance (ESR) spectroscopy possible to detect the radicals. After bombardment with heavy ions at about 100K, the ESR spectra of thymine contain an allyl radical (Structure 1), the octet pattern of the 5-thymyl radical (Structure 2) and contributions of the 6-yl radical (Structure 3) formed by net hydrogen gain at carbon $C_5$ [20]. The latter species (Structure 4 and 5) are also present in cytosine ( Structure 4 and 5) which in addition displays the pattern due to H-addition at $C_5$ and at the carboxyl oxygen, respectively. Adenosine exhibits two H-addition radicals (Structure 6 and 7), with the unpaired spin density located mainly at $N_1$ and $N_7$.

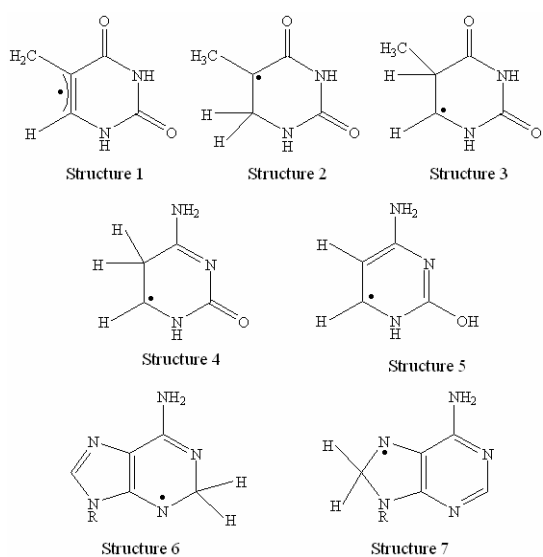

In discussing the possible mechanism of radical formation, solid-state ESR studies from nucleobase damage induced by sparsely ionizing irradiation may be taken into consideration [20,21]. It was suggested that those processes of radical formation can be interpreted as involving protonation of anionic and deprotonation of cationic states, which is called the "ionization pathway". Another possible mechanism proposed is an "excitation pathway" that results in homolytic cleavage producing a hydrogen radical. For example, the allyl radical (Structure 1) is formed by the loss of a hydrogen atom from the methyl group, leaving an unpaired electron in a $2p_z$ orbital of the methyl carbon atom. For these two mechanisms, although their exact assignment is often ambiguous, it was believed that the formation of most radicals in DNA and its constituents is mainly due to the "ionization pathway" [20].

Fragmentation is another primary step in ion-radiation induced damage of nucleobases. A high-energy ion can fragment biomolecules into pieces because of its high kinetic energy. Normally, gas phase studies in a well-defined experimental apparatus are preferred when a detailed molecular fragmentation mechanism is desired. To uncover the hyperfine structure of the fragment species, the Huels research group in Canada has investigated the heavy-ion induced fragmentation of fundamental DNA components in the gas phase, employing ions with very low energy (10-200 eV) [13,22-24]. The ions in this energy range are believed to be the primary ions at their track ends (the Bragg peak) or to be the secondary ions produced along the primary ion track. Deng et al. applied 200 eV $Ar^+$ ions to thymine (T) films and found numerous positive and negative ion fragments were produced via endocyclic and exocyclic bond cleavage [23,24]. The major cations desorbing from T films are identified as $HNCH^+$, $HN(CH)CCH_3^+$, $[T–OCN]^+$, $OCNH_2^+$, $C_xH_y^+$ (x=1-3 and y=0-4), $[T+H]^+$, and $[T-O]^+$ (Fig. 1(a)). Anion desorption is dominated by $H^-$, $O^-$, $CN^-$, $OCN^-$, and $[T–H]^-$, with lesser dissociation channels leading to desorption of $C^{2-}$, $C_2H^-$, $C_2CN^-$, $NC_3H_2^-$, $HNC_3H_3^-$, $OC_3H_3^-$, $C_2OCN^-$, and $C_3H_x^-$ (x=2 and 3). Figure 2 shows a possible reaction channel for the formation of $OCN^-$ and $[T-OCN]^+$. According to Blanksby and Ellison, the C-N bonds in a thymine molecule has energy roughly 4.5eV [25]. So the 200eV incident ion is powerful enough to beak them. Measurement of the dependence of fragment desorption yields upon the primary ion energy shows that the threshold energy for thymine fragmentation is very low (about 15 eV); the endocyclic and cationic fragments appear at much lower energies than exocyclic and anionic ones, respectively.

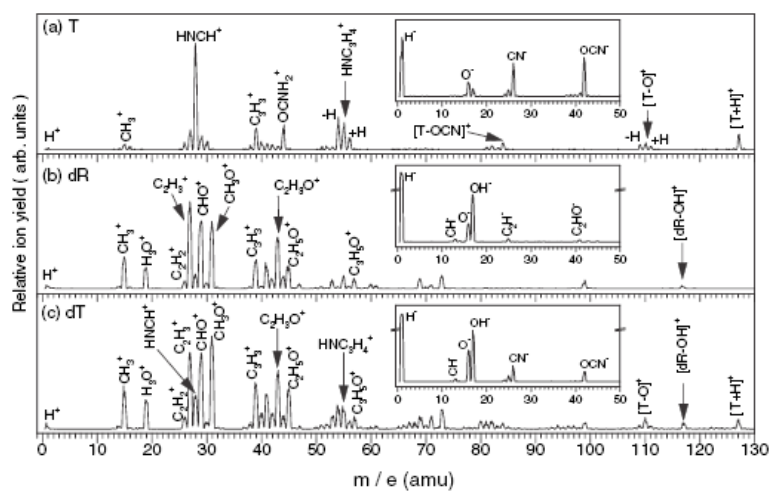

Fig.1. Cation and anion (insets) desorption mass spectra produced by 100 eV $Ar^+$ impact on films of (a) thymine (T), (b) 2-deoxy-D-ribose (dR), and (c) thymidine (dT). (Reprinted with kind permission from [13]. Copyright (2005) by the American Society of Physics).

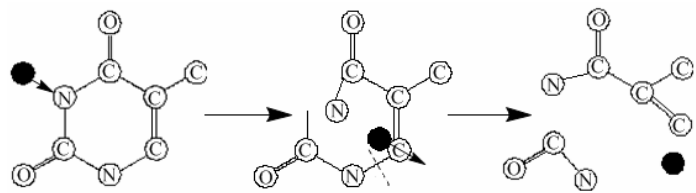

Fig.2 Schematic illustration of the formation of $OCNH_2^+$, $OCN^-$, and $[T-OCN]^+$ in the thymine backbone after an atomic collision cascade initiated by a low energy ion.

Multiply charged ions play an important role in biological radiation damage. Both theory [26] and experiments [27-29] have struggled to investigate the response of nucleobases upon interaction with multiply charged ions which may be one kind of secondary fragments along the radiation track of a primary heavy particle and may be able to cause substantial direct physical damage to DNA components. At the KVI in the Netherlands, Schlathölter and coworkers used an electron cyclotron resonance ion source to generate ion beams with charge states $q$ ranging from 1 to 25 at kinetic energies from $q \times 1$ keV to $q \times 25$ kev, which were then collimated and guided into the collision region to cross with gaseous targets of nucleobases [27,28]. The products were analyses by a time-of-flight (TOF) mass spectrometer. The results showed that, for keV carbon ions, the fragmentation dynamics are strongly influenced by the $C^{q+}$ charge state: for low $q$ values, fragmentation due to direct collisions is mainly observed; with increasing $q$, electron capture becomes more important; for larger $q$, several Coulomb explosion channels, which can lead to very energetic fragments have been identified [27,28]. In the case of highly charged ions (HCI), such as $Xe^{25+}$, the Coulomb explosion of nucleobases uracil and thymine can throw out singly and multiply charged atomic fragment ions $C^{q+}$, $N^{q+}$ and $O^{q+}$ ($q$ = 1-3) [29]. For the lowly charged ions, by comparing the spectra obtained through bombarding gaseous uracil with 1 keV/amu $He^{2+}$, $C^{2+}$, $N^{2+}$, and $O^{2+}$, it was found that a large fraction of the spectra for the four projectiles are quite similar, but the fragments initiated by $He^{2+}$ are richer than those by $C^{2+}$, $N^{2+}$, and $O^{2+}$ [28]. Other than the differences in electronic structure between the projectile ions, we would expect that this is due to the carbon and/or hydrogen abstraction reactions of the non-inert element ions which make their collision activities descend [22].

2.2. Nucleosides, nucleotides and oligonucleotides

A nucleotide is a chemical compound that consists of a heterocyclic base, a sugar, and one or more phosphate groups, while the nucleoside molecule includes only the former two species. As well as of thymine, Deng and his colleagues [13] have reported measurements of ion-molecule collision induced fragmentation of thymidine and 2-deoxy-D-ribose in the condensed phase by using their "soft-landed" ion beam apparatus. According to their experiments, nucleoside damage pathways include base or sugar loss, and complete disintegration of either moiety. As shown in Fig. 1, the fragments $[T+H]^+$ and $[dR-OH]^+$ are indicative of release of the whole base and sugar moieties, while the low-mass peaks imply the destruction of either ring. The fragmentation of solid-state thymidine and deoxyadenosine induced by slow multiply charged ions (200 keV $Xe^{10+}$ and 400 keV $Xe^{20+}$) has also been investigated by Manil and his colleagues [30]. In this case, larger molecules of modified dimers or trimers (m > 200) are produced in the sputtering process, as well as singly and doubly charged small-size fragments (m < 100). This result shows that ions with high charge states and high energies can more readily produce parent molecular ions and their crosslink ions than the low charged ones. The authors still specified the separation of the molecules into their component base and sugar parts as an important reaction channel.

However, this doesn't mean that the release of the whole base or sugar moiety is the major process of ion impact of nucleosides. On the contrary, from Fig. 1 it can be found that the disintegration of either moiety is more dominant by comparing the relative abundance of the low-mass fragment ions and that of the ringed ions.

It is well known that the furanose ring is puckered rather than planar. The furanose molecule has relatively high internal energy because of its twisted molecular structure. Otherwise, unlike the purine and pyrimidine molecules, the furanose molecule has no unsaturated bonds which can stabilize the molecule against external disturbance. Therefore, the sugar moiety in DNA should be more sensitive to ion attack than the nucleobase part, and the destruction of sugar appears to occur more readily. In 5 keV He$^+$ collision with gaseous adenine and deoxyribose, such differences have been observed [31]. The fragmentation patterns acquired from adenine [31] and other nucleobases [27,28] exhibit a bimodal distribution with decreasing intensities reaching a minimum (around group 7 for adenine) and increasing intensities afterwards up to the parent molecule (Fig. 3). In contrast to the nucleobases, deoxyribose shows a very small parent ion peak, and the overall peak intensities decrease monotonically with *m/z*. Moreover, the chemical identification and comparison of the fragments of thymine, 2-deoxy-D-ribose, and thymidine also show that in the thymidine molecule the majority of the low-mass fragments originate from the sugar moiety, but not from the base (see Fig. 1). These thoughtprovoking differences are thought to be related to the differential appearance energies (AE) of different fragments [31]. For example, for group 8 in Fig. 3(b), in addition to the 8.2 eV ionization energy, only 0.3 eV of extra energy is needed to form the fragment at *m/z* = 116 in deoxyribose, which is much lower than the 3.4 eV required for the fragment at *m/z* = 108 in adenine (Fig. 3(a)). Therefore, although some primary processes are understood, it is necessary to perform more thorough investigation on various DNA building blocks in order to fully fathom the ion-DNA interaction mechanism.

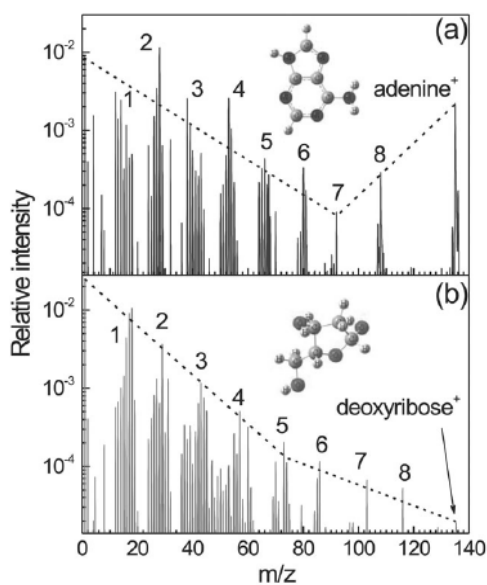

Fig. 3  Mass spectra of adenine (a) and deoxyribose (b) fragments after collisions with 5 keV amu$^{-1}$ He$^+$ ions. As an inset in each graph the structure of the correspondent molecule is shown. (Reprinted with kind permission from [31]. Copyright (2006) by the Royal Society of Chemistry).

Phosphate ester cleavage and release of inorganic phosphate (Pi) is another important characteristic of ionizing radiation of nucleotides and DNA [32,33], because reactions that cause release of Pi are equivalent to splitting the DNA strand and will inevitably give rise to single- and double-strand breaks. Shao and Yu have investigated the damage to nucleotides (e.g., 5'-dTMP, 5'-AMP, 5'-CMP) irradiated by $N^+$ ion beams with energies of 20-30 keV [34,35]. It was found that the main radiolysis reactions of ion irradiated nucleotides are Pi release and base release. The yield of free base is higher than that of inorganic phosphate, possibly indicating that the base is released first followed by the inorganic phosphate, just as Scholes et al. have suggested [36]. Studies with selective radical scavengers, histidine, implicate no OH radical has been formed in the solid state nucleotide samples. In the body of a living organism, however, owing to the existence of water, plenty of OH radical will be formed after ion bombardment, and the release of Pi will be more serious, because OH can react with the sugar moiety by abstracting hydrogen. The major product of H-abstraction in DNA is the $C_{4'}$ radical [37]. There now exists substantial evidence that the $C_{4'}$ radical undergoes a β-elimination of one of the phosphates and is a precursor to prompt the release of Pi (Fig. 4), followed by single strand breaks (SSBs) or double strand breaks (DSBs) [17,38,39].

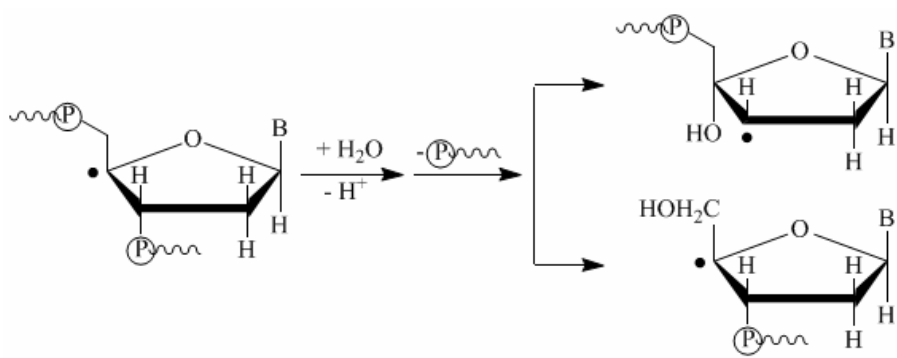

Fig. 4 The $C_{4'}$ radical mechanism for the release of inorganic phosphate in nucleotides and the formation of single strand break in DNA.

More information about fragmentation of ion-impacted DNA can also be obtained from their mass spectra, although the researchers have mainly focused on the analytical processes and have essentially ignored the intrinsic effects on the biological specimens. The low-energy gas atoms used in mass spectrometers are quite efficient in collision, causing ionization, excitation, and molecular fragmentation. Base loss and backbone fragmentation are the two major collision-induced dissociation (CID) events of RNA and DNA oligonucleotides [40]. The initial step of fragmentation is loss of nucleobase either as an anion or as a neutral species, and the overall anionic base loss follows the trend $A^- \gg G^- \approx T^- > C^-$ [41]. The GC base pair has three hydrogen bonds, whereas the AT base pair has only two, and as a consequence, the GC pair is more stable. However, it is puzzling that the CID experiments showed that between two DNA duplexes with the same melting temperature, the one with the higher GC content yields a higher fragment ion abundance, showing a higher radiolytic instability [42].

2.3. DNA

The analysis of ion impact with small DNA components is not strong enough to reveal the effects of ion irradiation on DNA. For a better understanding of the damage to genetic material induced by charged particles, the target material has been extended from simple building blocks to small plasmid molecules [43,44], genomic DNA of bacteria, and even of eukaryotes [45,46]. At the macromolecular level, the damage inflicted upon DNA by ion irradiation can be classified as either SSBs or DSBs. As the result of DNA components damage, the induction of these two lesions is an inevitable phenomenon due to several processes such as molecular ionization and excitation, direct collisions induced atom displacements, collision cascades of the constituent atoms, molecular bond-scissions and fragmentation (Fig. 2).

Experimental studies on the direct effects of ion irradiation in DNA molecules have been carried out using naked plasmid DNA. After irradiation, the production of SSBs and DSBs can be quantified by gel electrophoresis and image analysis software. By using these methods, the amount of DNA damage and its heterogeneity were found to increase with particle fluence and linear energy transfer (LET), but saturate at high fluences and LETs. This has been demonstrated by both *in vivo* and *in vitro* experiments [43,47]. For 30 keV $N^+$ irradiated pGEM3zf(+) DNA, the ratio of efficiencies of DSB and SSB production is about 6:1, indicating that production of DSB is much higher than that of SSB, and DNA fragments are the major form of damage initiated by nitrogen ion irradiation. Similarly, with even lower ion energy, for example, 1 keV $Ar^+$, SSBs and DSBs are also present in the pUC18 plasmid DNA [48]. About 14.5%, and 16.8% circular relaxed DNA (SSB) are detected after irradiation with a total of $1.2 \times 10^4$ and $2.4 \times 10^{14}$ ions, respectively, which are more than twice that of the vacuum control sample (~7%). But the DSB/SSB ratios after 1 keV $Ar^+$ irradiation of pUC18 plasmid are much lower than those of pGEM3zf(+) plasmid bombarded by keV $N^+$ [47,49], with values less than 1:10. For the fragmentation patterns of DNA molecules, there is no agreement on fragmentation sites. Based on analyzing the entire distribution of radiation-induced fragments [46], instead of only measuring the large fragments [45,48], it is concluded that high LET radiations are less effective than X-ray in producing large DNA fragments but more effective in the production of smaller fragments, supporting a non-random distribution of double-strand breaks induced by particle irradiation.

DNA structures can also be directly monitored by atomic force microscopy (AFM). AFM images revealed prodigious structural changes in the plasmid DNA due to HCI ($Xe^{44+}$) bombardment [50]. From Fig. 1(a), it is clear that most of the natural plasmid DNA are in a supercoiling structure or an open circle. After HCI irradiation the original circular structure of the DNA molecules is no longer recognizable. Some dense molecular clusters come into view, which may result from multiple DNA fragment aggregation. Hence, the ion implantation induced effects on the plasmid configuration intrinsically reflect the chemical laws of DNA atomic displacement, molecular rearrangement, and fragment crosslink. By using AFM, Brons et al compared the fragmentation patterns of ΦX-174 plasmids upon irradiation with heavy ions and X-rays [44]. According to their work, different from those obtained by similar dose X-ray irradiation, the fragment distributions resulting from 3.5 MeV $^{59}$Ni beams shift to smaller fragment sizes. Even with single ion traversals, the resulting average fragment length is significantly smaller than the full plasmid length. The exact fraction of the three main conformations of plasmid DNA can be distinguished on the basis of their different electrophoretic mobilities on agarose gels. According to the relative amounts of each formafter ion irradiation, the fluence or energy effects can be established and cross sections for SSB or DSB induction can be calculated.

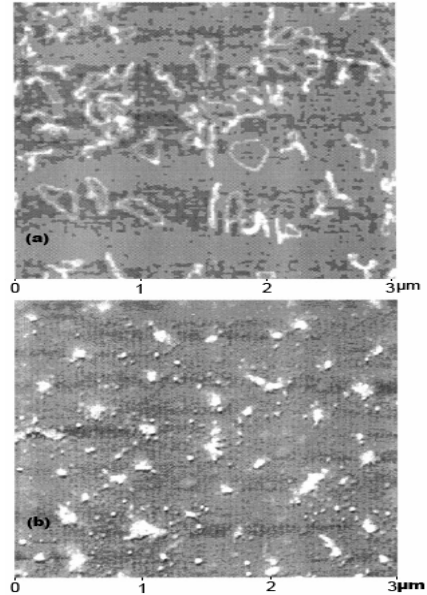

Fig. 5. AFM images of plasma DNA on mica. (a) before irradiation, showing intact circular plasmids; (b) after irradiation with Xe$^{44+}$ ions, showing profound damage. (Reprinted with permission from [50]. Copyright (1998) by Institute of Physics ).

In living organisms, most of the ionizing radiation induced biological lesions result from SSBs and DSBs. SSBs are usually repairable without further consequences, but DSBs often cause cell death or genetic mutations. Multiple close ionization events along the particle tracks, especially at the "Bragg peak", can form complex clusters of lesions which are even more difficult to repair and thus potentially more lethal or mutagenic [48,51]. Exactly owing to this, i.e., a controllable spatial distribution of stopping particles and a maximum dose density deposited at the "Bragg peak" shortly before the track end, ion beams have a high selectivity for tumor tissue and represent a promising radiotherapy modality for the treatment of deep-seated tumors [9,52]. On the other hand, DSBs constitute critical lesions which can lead to mutation if not correctly repaired. A good way to elucidate the mechanism is to irradiate the DNA *in vitro* and then transfer the DNA to host cells where the induced DNA lesions may be partially repaired and inherited via post-irradiation replication [53]. When the genetic substance DNA is damaged, the corresponding DNA repair enzymes are activated to induce repair. However, it is inevitable that some errors occur and these can lead to genetic mutations. This may be the foundation of ion-beam assisted radiation breeding.

**3. Epilogue**

Ion beams have been widely applied in a few biological research fields such as radioactive breeding, health protection, and tumor therapy. Up to now many interesting and impressive achievements in biology and agriculture have been made. To uncover the mechanisms of the radiobiological effects caused by ion irradiation, some chemical models in gas-, liquid-, and solid states have been employed to study the effects of ion irradiation on biomolecules, especially the genetic substance DNA. It has been suggested that the damage to the genome induced by sparsely ionizing radiation, such as X- and γ-rays, in a living cell by ionizing radiation is about two-thirds indirect and one-third direct [15,16].

In this paper, a chemical overview of the direct ion radiation effects on DNA and its components is presented, based on a review of literature combined with recent experimental results. It is suggested that, under ion bombardment, a DNA molecule undergoes a variety of processes, including radical formation, atomic displacement, intramolecular bond-scissions, emission of fragments, fragment recombination and molecular crosslink, which may lead to genetic mutations or cell death. These chemical and biological events make ion implantation a promising tool for radiation breeding and tumor therapy. Our goal with this contribution is to provide the reader with a perspective on this topic rather than a comprehensive review. Although many accomplishments have been achieved [1,2], the pertinent research still constitutes a weighty undertaking towards a distant destination.

**Acknowledgments**

The authors would like to express special thanks to the publishers (APS, RSC, and IOP) and the authors for their kind figure reprint permissions. We are grateful to Dr. H. J. Cleaves at CIW (Carnegie Institution of Washington ) for help in manuscript modification. Our thanks also go to those who have made concerted efforts to explore the chemical, physical and biological mechanisms of the radiobiological effects of ion beam.

**References**
[1] Z.L. Yu (*ed.*), L.D. Yu, T. Vilaithong, I. Brown (*trans.*), *Introduction to Ion Beam Biotechnology* (Springer, New York, 2006).
[2] H.Y. Feng, Z.L. Yu and P.K. Chu, Mat. Sci. Eng. R. **54** 49 (2007).
[3] L.F. Wu and Z.L. Yu, Radiat. Environ. Biophys. **40** 53 (2001).
[4] L.J. Wu, G. Randers-Pehrson, A. Xu *et al.*, Proc. Natl. Acad. Sci. USA **96** 4959 (1999).
[5] W. Han, L. Wu, S. Chen *et al.*, Oncogene, **26** 2330 (2007).
[6] J. Yang, X. Jing, W. Li *et al.*, Radiat. Environ. Biophys. **45** 261 (2006).
[7] X. Liu, K. Cai, H. Feng et al., Radiat. Environ. Biophys. **46** 255 (2007).
[8] D. Schulz-Ertner, O. Jäkel and W. Schlegel, Semin. Radiat. Oncol. **16** 249 (2006).
[9] D. Scholz-Ertner, A. Nikoghosyan, C. Thilmann *et al.*, Int. J. Radiat. Oncol. Biol. Phys. **58** 631 (2004).
[10] S. Anuntalabhochai, R. Chandej, B. Phanchaisri *et al.*, Appl. Phys. Lett. **78** 2393 (2001).
[11] Z. Tu, Y. Kobayashi, K. Kiguchi *et al.*, Nucl. Instrum. Meth. B **206** 591 (2003).
[12] Z.L. Yu, IEEE Transact. Plasma Sci. **28** 128 (2000).
[13] Z.W. Deng, I. Bald, E. Illenberger *et al.*, Phys. Rev. Lett. **95** 153201 (2005).
[14] Q. Li, Y. Furusawa, M. Kanazawa *et al.*, Nucl. Instrum. Meth. B **245** 302 (2006).
[15] C. von Sonntag, *The Chemical Basis of Radiation Biology* (Taylor and Francis, London, 1987).
[16] D. Michael and P. O'Neill, Science **287** 1603 (2000).
[17] P. O'Neil, *Radiation-induced damage in DNA*. In: Jonah CD, Rao BSM (*eds.*), *Radiation Chemistry: Present Status and Future Trends*. pp585-622 (Elservier, Amsterdam, 2001).
[18] W. Friedland, P. Bernhardt, P. Jacob *et al.*, Radiat. Prot. Dosim. **99** 99 (2002).
[19] J.G. Deng, Z.L. Yu and L.J. Qiu, Nucl. Tech. **15** 601 (1992).
[20] A. Schaefer, J. Hüttermann and G. Kraft, Int. J. Radiat. Biol. **63(2)** 139 (1993).
[21] W.A. Bernhard, Adv. Radiat. Biol. **9** 199 (1981).
[22] Z. Deng, I. Bald, E. Illenberger E *et al.,* Phys. Rev. Lett. **96** 243203 (2006).
[23] Z. Deng, M. Imhoff and M.A. Huels, J Chem. Phys. **123** 144509 (2005).
[24] M. Imhoff, Z. Deng and M.A. Huels, Int. J. Mass. Spectrom. **245** 68 (2005).
[25] S.J. Blanksby and G.B. Ellison, Acc. Chem. Res. **36** 255 (2003).
[27] J. de Vries, R. Hoekstra, R. Morgenstern *et al.*, Eur. Phys. J. D **24** 161 (2003).

[28] T. Schlathölter, F. Alvarado and R. Hoekstra, Nucl. Instrum. Meth. B **233** 62 (2005).
[29] J. de Vries, R. Hoekstra, R. Morgenstern *et al.*, Phys. Rev. Lett. **91** 053401 (2003).
[30] B. Manil, H. Lebius, B.A. Huber *et al.*, Nucl. Instrum. Meth. B **205** 666 (2003).
[31] F. Alvarado, S. Bari, R. Hoekstra *et al.*, Phys. Chem. Chem. Phys. **8** 1922 (2006).
[32] J.A. Raleigh, C.L. Greenstock and W. Kremers, Int. J. Radiat. Biol. **23** 457 (1973).
[33] J.F. Ward and I. Kuo, Int. J. Radiat. Biol. **23** 543 (1973).
[34] C.L. Shao and Z.L. Yu, Radiat. Phys. Chem., **44** 651 (1994).
[35] C.L. Shao, X.Q. Wang and Z.L. Yu, Radiat. Phys. Chem. **50** 561 (1997).
[36] G. Scholes, J.F. Ward and J. Weiss, J. Mol. Biol. **2** 379 (1960).
[37] M. Kuwabara, Z.Y. Zhang and G. Yoshii, Int. J. Radiat. Biol. **41** 241(1982).
[38] S. Steenken, G. Behrens and D. Schulte-Frohlinde, Int. J. Radiat. Biol. **25** 205 (1974).
[39] M. Dizdaroglu, C. von Sonntag, and D. Schulte-Frohlinde, J Am. Chem .Soc. **97** 2277 (1975).
[40] S. Schürch, E. Bernal-Méndez and C.J. Leumann, J Am. Soc. Mass Spectrom. **13** 936 (2002).
[41] S. Pan, K. Verhoeven and J.K. Lee, J Am. Soc. Mass Spectrom. **16** 1853 (2005).
[42] S. Pan, X. Sun and J.K. Lee, J Am. Soc. Mass Spectrom. **17** 1383 (2006).
[43] Y. Zhao, Z. Tan, Y. Du *et al.*, Nucl. Instrum. Meth. B **211** 211(2003).
[44] S. Brons, K. Psonka, M. Heiβ *et al.*, Radioth. Oncol. **73** S112 (2004).
[45] M. Lobrich, B. Rydberg and P.K. Cooper, Radiat. Res. 139(2) 142 (1994).
[46] M. Lobrich, P.K. Cooper and B. Rydberg. Int. J. Radiat. Biol. **70** 493 (1996).
[47] I. Testard and L. Sabatier, Mutat. Res. **448** 105 (2002).
[48] B. Rydberg, M. Lobrich and P.K. Cooper, Radiat. Res. **139(2)** 133 (1994).
[49] Y. Chen, B.Y. Jiang, Y. Chen *et al.*, Radiat. Environ. Biophys. **37(2)** 101 (1998).
[50] C. Ruehlicke, D. Schneider, M. Schneider *et al.*, Nanotechnology **9** 251 (1998).
[51] B.M. Sutherland, P.V. Bennett, O. Sidorkina *et al.*, Proc. Natl. Acad. Sci. USA **97** 103 (2000).
[52] M. Scholz, Heavy ion tumor therapy. Nucl. Instru. Meth. B **161-163** 76 (2000).
[53] Q. Wang, G. Zhang, Y.H. Du *et al.*, Mutat. Res.: Fund. Mol. M **528** 55 (2003).